\documentclass[11pt,twoside]{article}
\usepackage{asp2004}
\usepackage{lscape}
\usepackage{natbib}

\def\edcomment#1{\iffalse\marginpar{\raggedright\sl#1\/}\else\relax\fi}
\reversemarginpar

\begin{document}
\title{Type Ia Supernova Diversity from 3-dimensional Models}
\author{F. K. R{\"o}pke, M. Gieseler, and W. Hillebrandt}
\affil{Max-Planck-Institut f{\"ur} Astrophysik,
  Karl-Schwarzschild-Str.~1,\\85741 Garching, Germany}

\begin{abstract}
We present results from a systematic study of the effects of initial
parameters on three-dimensional thermonuclear supernova models.
\end{abstract}
\thispagestyle{plain}

One major challenge to numerical models of thermonuclear
supernova explosions is to reproduce the diversity observed among Type
Ia supernovae (SNe Ia). With the rapid advancement of
three-di\-men\-sion\-al models 
over the past years, systematic studies have become possible testing
the effect of changes in the initial parameters of simulations on the
explosion process. The ultimate goal is, of course, to explain the
correlation between the peak luminosity and the light curve shape
used to calibrate cosmological distance measurements.

We present the first systematic study on this issue based on
three-di\-men\-sion\-al deflagration models. Here, a Chandrasekhar-mass
white dwarf (WD) star ignites nuclear reactions in
the center which finally form a flame. In the deflagration model this
flame burns outward with a subsonic velocity, which quickly
becomes dominated by the effects of interaction with a turbulent
velocity field. Caused by buoyancy (Rayleigh-Taylor) and secondary
shear (Kelvin-Helmholtz) instabilities inherent in the scenario,
turbulence wrinkles the surface of the flame effectively boosting its
propagation speed. For
a comprehensive review of SN Ia models we refer to
\citet{hillebrandt2000a}.

For the three-dimensional simulations of the explosion process we
apply the scheme developed by \citet{reinecke1999a} and \citet{reinecke2002a}. The
final composition of the ejecta in our models is obtained via a
nucleosynthetic postprocessing procedure on the basis of tracer
particles advected in the explosion
simulation \citep[for details see][]{travaglio2004a}.

There is a number of possible parameters of the deflagration model
that have the potential to explain the SN Ia diversity. 
The progenitor's carbon-to-oxygen (C/O) ratio, its metallicity,
and the central density at ignition are commonly suggested, but other
effects like rotation and flame ignition could also play a role. In our
survey we concentrate on the former three parameters. These are varied
independently. Of
course, this is an oversimplification since in reality they are
interrelated by stellar evolution of the progenitor
WD star. Nevertheless, in this first study our goal is to infer the trends of
effects of each individual parameter on the explosion.

In the setup of the explosion models we apply three different carbon mass fractions of the WD
material, $X(^{12}\mathrm{C}) = 0.30, 0.46, 0.62$, and three different
central densities at ignition, $\rho_c = [1.0, 2.6, 4.2] \times 10^9
\, \mathrm{g} \, \mathrm{cm}^{-3}$. Three different metallicities of
the WD, $Z = [0.3, 1.0,
3.0] Z_\odot$, are represented by the
$^{22}$Ne mass fractions \citep[see][]{timmes2003a}. This defines the 27 models of our survey.

A change in the
carbon mass fraction does not alter the flame evolution in our models
significantly. This is due to a larger amount of $\alpha$-particles in
the nuclear statistical equilibrium in the ashes for carbon-rich fuel,
which act as an energy buffer \citep[see][]{roepke2004a}. This effect
delays the flame propagation at explosion stages where iron group
nuclei are synthesized and therefore the changes in the resulting
$^{56}$Ni masses are little. Since the radioactive decay of this
isotope powers the lightcurve, according to Arnett's rule the peak
luminosity is nearly unaffected by the C/O ration of the
progenitor. However, in the later evolution the
carbon-rich fuel models release more energy (amounting to a $\sim$12\%
variation) which possibly affects the
light curve shape.

A lower central density leads to a lower energy
release and delays the evolution of the model. The reason for this effect is the
reduced gravitational acceleration experienced the flame front which leads to
a slower formation of the nonlinear Rayleigh-Taylor
instabilities. We find a $\sim$40\% variation in the energy releases of
our models.
The amount of produced $^{56}$Ni is largest for
the intermediate central density. At lower densities the delayed flame propagation
leads to a decreased production of iron group elements.
In models with higher central densities electron captures favor
neutron rich nuclei instead of $^{56}$Ni.
This effect is accounted for
in the nucleosynthetic postprocessing but not consistently modeled in
the explosion simulation. Thus the results are preliminary for the
highest central density value. The change in the produced $^{56}$Ni is
$\sim$10\%. A variation in the central density prior to ignition
is likely to affect both the peak luminosity and the light curve shape.

Changing the metallicity (i.e.\ the $^{22}$Ne mass fraction) we find a
$\sim$20\% change in the resulting $^{56}$Ni masses and hence the peak luminosity
of the event consistent with the analytic
prediction of \citet{timmes2003a}. The explosion dynamics, however, is
not affected by the metallicity and thus the light curve shape will
not differ significantly.

Concluding we note that the variations in the peak
luminosities obtained by changing the initial parameters can partially
account for of the observed scatter in SNe Ia. However, we could not
identify a single parameter that reproduces the empirically
established peak luminosity--light curve shape relation. A combination
of initial parameters chosen according to stellar evolution could
possibly explain such a relation, but initial parameters ignored in
this first survey have to be explored as well in future studies.
The final decision on the effect of the initial parameters on the light curve
shape can only be made on the basis of synthetic light curve
calculations.

\acknowledgements{We thank M.~Reinecke and C.~Travaglio for their support
in this work.}


\end{document}